\documentclass[3p]{elsarticle}
\usepackage{amsmath,amsthm}
\usepackage{color}

\newtheorem{definition}{Definition}
\newtheorem{theorem}{Theorem}

\newcommand{\probp}{\ensuremath{\mathcal{P}}}
\newcommand{\alga}{\ensuremath{\mathcal{A}}}
\newcommand{\algb}{\ensuremath{\mathcal{A'}}}

\renewcommand{\times}{\cdot}

\begin{document}
\begin{frontmatter}
\title{A note on ``Approximation schemes for a subclass of subset selection problems'', and a faster FPTAS for the Minimum Knapsack Problem}

\author[cnam]{C\'edric Bentz}
\ead{cedric.bentz@cnam.fr}
\address[cnam]{CEDRIC-CNAM, 292 rue Saint-Martin, 75141 Paris Cedex, France}

\author[monash]{Pierre Le Bodic\corref{cor2}}
\ead{pierre.lebodic@monash.edu}
\address[monash]{Faculty of Information Technology, Monash University, Australia}

\cortext[cor2]{Corresponding author}

\begin{abstract}
 Pruhs and Woeginger \cite{pruhs07a} prove the existence of FPTAS's for a general class of minimization and maximization subset selection problems.
 Without losing generality from the original framework, we prove how better asymptotic worst-case running times can be achieved if a $\rho$-approximation algorithm is available, and in particular we obtain matching running times between maximization and minimization subset selection problems.
 We directly apply this result to the Minimum Knapsack Problem, for which the original framework yields an FPTAS with running time $O(n^5/\epsilon)$, where $\epsilon$ is the required accuracy and $n$ is the number of items, and obtain an FPTAS with running time $O(n^3/\epsilon)$, thus improving the running time by a quadratic factor in the worst case.
\end{abstract}

\begin{keyword}
Approximation algorithm \sep Approximation scheme \sep FPTAS \sep Worst case analysis \sep Pseudo-polynomial algorithm \sep Combinatorial optimization \sep Minimum Knapsack Problem
\end{keyword}

\end{frontmatter}

 \section{Introduction}
 Pruhs and Woeginger \cite{pruhs07a} give a simple and elegant proof of the existence of an FPTAS for a set of general subset selection problems, provided there exists an exact algorithm that runs in pseudo-polynomial time.
 We recall the main definition and result.
 \begin{definition}[\cite{pruhs07a}]
 A subset selection problem \probp{} is a combinatorial optimization problem whose instances $I = (X, w, S)$ consist of:
 \begin{itemize}
  \item a ground set $X$ with $\vert X \vert = n$ elements;
  \item a positive integer weight $w(x)$ for every $x \in X$;
  \item a structure $S$ that is described by $\ell(S)$ bits.
 \end{itemize}
 \end{definition}
 The structure $S$ controls which subset of elements of $X$ is feasible or infeasible, and its encoding length $\ell(S)$ is used when analyzing the running time of an algorithm.
 As in \cite[Chapter 1]{garey79}, we assume that $\ell(S)$ reflects the encoding length of the structure $S$ in a ``reasonable encoding scheme'' of $I$.
 An example is given in Section \ref{sec:example on min kp}.
 The sum of the weights of an instance $I = (X, w, S)$ will be referred to as the ``total weight'' and is equal to $W := \sum_{x \in X} w(x)$.
 In all the problems that will be considered in this note, one looks for a solution, i.e., one wishes to select a subset $X' \subseteq X$, which is \emph{feasible} (i.e., respects some constraint given by $S$) and either maximizes or minimizes the quantity $\sum_{x \in X'} w(x)$.

Examples of such problems include the Maximum Independent Set Problem (a maximization problem where elements are vertices of a graph, and the structure $S$ ensures that only non-adjacent elements can be selected), and the Maximum Knapsack Problem (a maximization problem over a set of items, the elements, where the structure $S$ contains a single packing constraint). Note that, among these problems, some are strongly NP-hard (like the former one \cite{garey79}), while others are weakly NP-hard (like the latter one \cite{garey79}).

 Throughout the paper, we use polynomials of the form $p(n, M, \ell(S))$ to bound asymptotic running times of algorithms, where $M$ is a value that depends on $n$ and/or $W$, and can for instance be equal to $W$.
 Since the first and the third arguments are invariably $n$ and $\ell(S)$, respectively, we will simply use the notation $p(M)$ (where $M$ is substituted for the \textit{ad hoc} value) wherever clarity is unimpaired.

 \begin{theorem}[\cite{pruhs07a}]\label{th:fptas without bound}
  Let \probp{} be a subset selection problem with instances $I = (X, w, S)$ that satisfies the following condition:\\
  $(C)$ There exists an algorithm \alga{} that solves \probp{} to optimality and whose running time is bounded by a polynomial $p$ in $n$, in total weight $W$, and in $\ell(S)$.\\
  Then problem \probp{} has an FPTAS with running time $n \times p(n^3/\epsilon)$ for minimization and $n \times p(n) + p(n^2/\epsilon)$ for maximization.
 \end{theorem}

 We refer to \cite{pruhs07a} for all other definitions, notation and references.

 In their paper, Pruhs and Woeginger point out that many existing proofs for the existence of FPTAS's for subset selection problems require a ``separate argument'' to provide a lower or upper bound (respectively, for minimization or maximization problems) on the optimal value, and they provide a graceful proof that does not use bounds.
 However, the resulting FPTAS may have a higher asymptotic running time that one that uses a lower or upper bound.
 The contribution of this note is Theorem \ref{th:fptas with bound}.
 \begin{theorem}\label{th:fptas with bound}
  Let \probp{} be a subset selection problem with instances $I = (X, w, S)$ that satisfies $(C)$.
 Suppose w.l.o.g. that the following condition holds:\\
 $(C')$ There exists a $\rho$-approximation algorithm $\algb$ to problem \probp{} whose running time is bounded by a polynomial $p'$ in $n$, in $M =\log(W)$, and in $\ell(S)$.

  Then problem \probp{} has an FPTAS with running time $p'(\log(W)) + p(n^2/\epsilon)$ (for both minimization \emph{and} maximization), if $\rho$ is a constant.
 \end{theorem}
 In Theorem \ref{th:fptas with bound}, we can suppose without loss of generality that a $\rho$-approximation algorithm \algb{} exists, since \probp{} satisfies $(C)$, and thus there exists an FPTAS by Theorem \ref{th:fptas without bound}.
 The proof of Theorem \ref{th:fptas with bound} is given in Section \ref{sec:proof}, and the running time analysis in Section \ref{sec:time}.
 Although it is possible to use \alga{} with a given $\epsilon$ instead of algorithm \algb{} in Theorem \ref{th:fptas with bound}, doing so may not provide running time improvements.
 However, we will prove in Section \ref{sec:example on min kp} that Theorem \ref{th:fptas with bound} provides an FPTAS with a faster running time than Theorem \ref{th:fptas without bound} on an important example, the Minimum Knapsack Problem.

 \section{Proof of Theorem \ref{th:fptas with bound}}\label{sec:proof}
 The proof follows the one given by Pruhs and Woeginger in \cite{pruhs07a}, with some simplifications due to assumption $(C')$.
 To begin with, we will not need to suppose that items are ordered by non-decreasing weights.

 \subsection{Proof for minimization problems}
 Assumption $(C')$ implies that, for a given $\rho > 1$, and for any instance $I$,
 \begin{align} \label{eq: UB vs OPT}
 OPT \leq UB \leq \rho \times OPT,
 \end{align}
 where $OPT$ is the optimal value of instance $I$ and $UB$ is the upper bound returned by \algb{} on $I$.

 We define the scaling parameter
 \begin{align}\label{eq: def of scaling Z for min}
  Z := \frac{\epsilon}{\rho} \times\frac{1}{n} \times UB.
 \end{align}

We create a modified instance $I'$ where only weights differ from $I$.
Let $w'$ be the weights of $I'$ such that, for $i=1, \dots, n$,
\begin{align} \label{eq:def of w' for min}
 w'(x_i) =
 \begin{cases}
 \lceil w(x_i)/Z \rceil &\text{if $w(x_i) \leq UB$,} \\
 UB/Z + n + 1 & \text{otherwise.} \\
 \end{cases}
\end{align}

The total weight $W'$ of instance $I'$ is thus bounded by
\begin{align}
 W' \leq
 n \times (UB/Z + n + 1) \leq
 n \times (\rho \times n / \epsilon + n + 1) . \label{eq: bound on W' for min}
\end{align}
Since $\rho$ is fixed, instance $I'$ can be solved in a time polynomial in the size of instance $I$ and in $1/\epsilon$ by \alga{}.
Let $Y^*$ denote an optimal solution (of value $OPT$) for instance $I$, and $Y'$ for instance $I'$.
Observe that, because $OPT \leq UB$, any item $i$ in $Y^*$ satisfies $w(x_i) \leq UB$.
Since $Y'$ is an optimal solution of $I'$, and $Y^*$ is feasible for $I'$ (since the structure $S$ is the same in $I$ and in $I'$), we have:
\begin{subequations}
\begin{align}
\sum\{w'(y) \vert y \in Y'\}
&\leq \sum\{w'(y) \vert y \in Y^*\} \label{eq: opt vs feas for min} \\
&\leq \sum \{ \lceil w(y) / Z \rceil \vert y \in Y^*\}\\
&\leq \sum \{ w(y) / Z +1 \vert y \in Y^*\}\\
&\leq OPT/Z + n \label{eq: shortcut for analysis}\\
&\leq UB/Z + n. \label{eq: no big item}
\end{align}
\end{subequations}
Equations \eqref{eq: no big item} and \eqref{eq:def of w' for min} imply that, if there exists $i \in \{1, \dots, n\}$ such that $w(x_i) > UB$, $x_i$ cannot belong to an optimal solution $Y'$ to $I'$.
We are now ready to prove that $Y'$ is a $(1+\epsilon)$-approximate solution to $I$.
\begin{subequations}
\label{eq:analysis for min}
\begin{align}
 \sum\{w(y) \vert y \in Y'\}
 & \leq Z \times \sum \{w'(y) \vert y \in Y'\} && \text{by } \eqref{eq:def of w' for min}\\
 & \leq OPT + n \times Z && \text{by } \eqref{eq: shortcut for analysis}\\
 & = OPT + n \times \frac{\epsilon}{\rho} \times\frac{1}{n} \times UB && \text{by } \eqref{eq: def of scaling Z for min} \\
 & \leq ( 1 + \epsilon ) \times OPT && \text{by } \eqref{eq: UB vs OPT}.
\end{align}
\end{subequations}

\subsection{Proof for maximization problems}
 For maximization problems, assumption $(C')$ implies that, for a given $\rho < 1$,
 \begin{align} \label{eq: LB vs OPT}
 OPT \geq LB \geq \rho \times OPT,
 \end{align}
where $LB$ is the value returned by \algb{}.

We define the scaling parameter $Z$ as
 \begin{align}\label{eq: def of scaling Z for max}
  Z := \epsilon \times\frac{1}{n} \times LB.
 \end{align}
We create a modified instance $I'$ where only weights differ from $I$.
We use the property that, if $w(x_i) > LB /\rho$ for some $i \in \{1, \dots, n\}$, then item  $x_i$ cannot belong to a feasible solution to $I$ (and hence to $I'$), because of \eqref{eq: LB vs OPT}.
Therefore, we encode the weights $w'$ of $I'$ as
\begin{align} \label{eq:def of w' for max}
 w'(x_i) =
 \begin{cases}
 \lfloor w(x_i)/Z \rfloor &\text{if } w(x_i) \leq LB / \rho, \\
 1 & \text{otherwise. }\\
 \end{cases}
\end{align}
The total weight $W'$ of instance $I'$ is thus bounded by
\begin{align}
 W'
 &\leq \sum_{i=1}^{j-1} \left\lfloor \frac{n \times w(x_i)}{\epsilon \times LB} \right\rfloor + \sum_{i=j}^{n} 1 \text{ (for some $j \geq 2$)} && \text{by } \eqref{eq: def of scaling Z for max} \nonumber\\
 & \leq n \left\lfloor \frac{n}{\epsilon \times \rho} \right\rfloor + n . \label{eq: bound on W' for max}
\end{align}
Since $\rho$ is fixed, instance $I'$ can be solved in a time polynomial in the size of instance $I$ and in $1/\epsilon$ by \alga{}.
We use again notation $Y^*$ and $Y'$.
Because $Y^*$ is feasible for $I'$ (since the structure $S$ is the same in $I$ and in $I'$), we have
\begin{align}\label{eq: opt vs feas for max}
 \sum \{w'(y) \vert y \in Y'\} \geq  \sum \{w'(y) \vert y \in Y^*\}.
\end{align}

We now prove that $Y'$ is a $(1-\epsilon)$-approximate solution to $I$.
\begin{align*}
 \sum\{w(y) \vert y \in Y'\}
 & \geq Z \times \sum \{w'(y) \vert y \in Y'\} && \text{by } \eqref{eq:def of w' for max}\\
 & \geq Z \times \sum \{w'(y) \vert y \in Y^*\} && \text{by } \eqref{eq: opt vs feas for max}\\
 & = Z \times \sum \{ \lfloor w(y) / Z \rfloor \vert y \in Y^*\} && \text{by } \eqref{eq:def of w' for max} \\
 & \geq Z \times \sum \{ w(y) / Z -1 \vert y \in Y^*\}\\
 & \geq \sum \{ w(y) \vert y \in Y^*\} - \vert Y^* \vert \times Z\\
 & \geq OPT - n \times \epsilon \times\frac{1}{n} \times LB && \text{by } \eqref{eq: def of scaling Z for max} \\
 & \geq ( 1 - \epsilon ) OPT && \text{by } \eqref{eq: LB vs OPT}.
\end{align*}

\section{Running time comparisons}\label{sec:time}
The following table gives the asymptotic worst-case running times if one only uses \alga{}, as in \cite{pruhs07a}, or both \alga{} and \algb{}, as in this paper.
\begin{center}
 \begin{tabular}{c|cc}
  & Minimization & Maximization \\
  \hline
  \alga{} & $n \times p(n^3/\epsilon)$  & $n \times p(n) + p(n^2/\epsilon)$ \\
  \hline
  \alga{} \& \algb{} & $p'(\log(W)) + p(n^2/\epsilon)$ & $p'(\log(W)) + p(n^2/\epsilon)$\\
\end{tabular}
\end{center}

 The FPTAS for minimization problems that uses \alga{} (see \cite{pruhs07a}) first sorts the items in $O(n \log n)$ time, constructs auxiliary instances in $O(n^2)$ time, and solves all instances in $O(n \times p(n^3/\epsilon))$ time.
 The FPTAS for maximization problems that uses \alga{} (see \cite{pruhs07a}) first sorts the items in $O(n \log n)$ time, solves an instance for each item in $O(n \times p(n))$ time, constructs an auxiliary instance in $O(n)$ time, and solves it in $O(p(n^2/\epsilon))$ time.
  The FPTAS for minimization or maximization problems that uses \alga{} and \algb{} builds an approximate solution in $p'(\log(W))$ time, constructs instance $I'$ in $O(n)$ time, and solves it in $O(p(n^2 / \epsilon))$ time (since $W'=O(n^2/\epsilon)$ if $\rho$ is fixed, from \eqref{eq: bound on W' for min} and \eqref{eq: bound on W' for max}).
In Section \ref{sec:example on min kp}, we show in particular how a quadratic improvement on the running time can be achieved on the Minimum Knapsack Problem using both \alga{} and \algb{}.

\section{Application of Theorem \ref{th:fptas with bound} to the Minimum Knapsack Problem} \label{sec:example on min kp}
The Minimum Knapsack Problem (MinKP) \cite[Chapter 13]{kellerer04a} has been studied less extensively than its maximization counterpart, MaxKP.
Basically, MinKP is a minimization problem over a set of items (the elements) where the structure $S$ contains a single covering constraint.
Constant-ratio approximation algorithms for this problem are given in \cite{carnes08a, csirik91a, guntzer00}, but, excluding \cite{pruhs07a}, there does not seem to be any \emph{explicit} description and/or analysis of an FPTAS for it in any reference books or articles.
The classical papers by Ibarra and Kim \cite{ibarra75a} and Lawler \cite{lawler79a} only deal with MaxKP, and the results in these papers do not seem to be directly applicable to MinKP.
Gens and Levner \cite{gens79a, gens80a} state that FPTAS's for MinKP exist, based in part on a paper by Sahni \cite{sahni75a}, but do not give proofs and only partial analyses.
The idea that the result of Pruhs and Woeginger \cite{pruhs07a} could be improved stemmed from Gens and Levner's papers.

We show how directly applying Theorem \ref{th:fptas with bound} to the Minimum Knapsack Problem yields a quadratic improvement over Theorem \ref{th:fptas without bound}.
Note that the FPTAS's these two theorems provide have the same asymptotic running time for the Maximum Knapsack Problem.

The structure $S$ that encodes (in)feasibility corresponds to the coefficients of the knapsack constraint, together with the right hand side of the constraint.
Its encoding length $\ell(S)$ hence corresponds to the sum of the binary encodings of each of these numbers.

The dynamic programming algorithm in e.g. \cite[Chapter 8]{vazirani01a} runs in $p(n, W, \ell(S))=n \times W$, and may be used as algorithm \alga{}.
(Note that, for exact resolution, minimum and maximum knapsack problems are equivalent, and hence both are weakly NP-hard \cite{garey79}.)
Using only algorithm \alga{}, Theorem \ref{th:fptas without bound} ensures the existence of an FPTAS that runs in $O(n^5/\epsilon)$.
The $2$-approximation given in \cite{csirik91a} runs in a time bounded by $p'(n, M = \log(W), \ell(S))= n^2$ and may serve as algorithm \algb{}.
Using both \alga{} and \algb{}, Theorem \ref{th:fptas with bound} guarantees the existence of an FPTAS with running time $O(n^3/\epsilon)$, which is a quadratic improvement over the FPTAS provided by Theorem \ref{th:fptas without bound}.

\section*{Acknowledgements}
The authors would like to thank Ulrich Pferschy for introducing them to the original paper \cite{pruhs07a}.

\section*{References}

\bibliography{biblio}
\bibliographystyle{plain}
\end{document}